\documentclass[twocolumn]{revtex4}
\usepackage[pdftex]{graphicx}
\usepackage{amsmath}
\usepackage{hyperref}
\usepackage[latin1]{inputenc}
\usepackage{vmargin}
\usepackage[english]{babel}
\usepackage{graphicx}
\usepackage[dvips]{epsfig}
\usepackage{amssymb}
\usepackage{bm}
\usepackage[sort&compress]{natbib}

\newcommand{\Alumina}{Al$_2$O$_3$}
\newcommand{\angst}{\,\mathrm{\AA}}
\newcommand{\ba}[1]{\begin{array}{#1}}
\newcommand{\bc}{\begin{center}}
\newcommand{\bea}{\begin{eqnarray}}
\newcommand{\be}{\begin{equation}}
\newcommand{\ea}{\end{array}}
\newcommand{\ec}{\end{center}}
\newcommand{\eea}{\end{eqnarray}}
\newcommand{\ee}{\end{equation}}

\newcommand{\fig}[1]{Fig.\,\ref{#1}}

\newcommand{\microns}{\mu\mathrm{m}}
\newcommand{\mmoll}{\,\mathrm{mmol/l}}
\newcommand{\pH}{p\mathrm{H}}
\newcommand{\kB}{k_{\mathrm{B}}}
\newcommand{\pHvalue}{\mbox{$\pH$-value}}
\newcommand{\Ref}[1]{Ref.~\cite{#1}}
\newcommand{\Refs}[1]{Refs.~\cite{#1}}

\renewcommand{\epsilon}{\varepsilon}
\renewcommand{\theta}{\vartheta}
\renewcommand{\vec}[1]{\mathbf{#1}}

\newcommand{\etal}{\,{\it et al.}\,}

\graphicspath{{}{images/}{submission/}}

\begin{document}

\title{Computational Steering of Cluster Formation in Brownian Suspensions}
\author{Martin Hecht}
\author{Jens Harting}
\affiliation{Institut f\"ur Computerphysik, Pfaffenwaldring 27, 70569
Stuttgart, Germany}
\date{\today}

\begin{abstract}
We simulate cluster formation of model colloidal particles interacting via DLVO 
(Derjaguin, Landau, Vervey, Overbeek) potentials.
The interaction potentials can be related to experimental conditions, defined by 
the \pHvalue{}, the salt concentration and the volume fraction of solid particles
suspended in water. The system shows different structural properties for different conditions,
including cluster formation, a glass-like repulsive structure, or a liquid suspension.
Since many simulations are needed to explore the whole parameter space, 
when investigating the properties of the suspension depending on the experimental conditions,
we have developed
a steering approach to control a running simulation and to detect interesting transitions
from one region in the configuration space to another. The advantages of the steering
approach and the restrictions of its applicability due to physical constraints are 
illustrated by several example cases.
\end{abstract}

\maketitle

\newlength{\epswidth}
\setlength{\epswidth}{\linewidth}

\section{Introduction}
Soft matter physics is a large field which has gained more and more in importance 
during the last years. It comprises for example complex fluids, biological systems like 
membranes, solutions of large molecules like proteins, or suspensions of small soft or solid 
particles, which are commonly called ``colloids''. Since one can find examples
for these materials nearly everywhere in everyday life (in medicine, food industry,
paintings, glue, blood, ceramics,\dots), many results in research have a high relevance 
for applications. In this context, especially colloid science is one of the most 
active research fields of the present time. A considerable effort has been 
invested to describe colloidal suspensions from a theoretical point of view and by 
simulations~\cite{Ladd94,Ladd94b,Brady96,Silbert97a,Harnau04} as well as to understand 
the particle-particle interactions~\cite{DLVO,DLVO2,Alexander84,Grimson91,vanRoij97,Dobnikar03} and
the phase behavior~\cite{Trappe01,Levin03,Costa04,Hynninen04,Sator04}. 
Typical for soft matter physics, and especially for colloids,
is that the important effects which govern the material properties are on a mesoscopic
length scale, i.e., on lengths much larger than the atomistic scale, and much smaller 
than the macroscopic length scale. 

This mesoscopic level brings in the possibility to control the parameters of a
certain material by manipulating the processes on the mesoscopic length scale. 
For colloids there are many ways to control the interactions between the individual
colloidal particles. As an example, by changing the particle interactions, it is 
possible to initiate a clustering process in a colloidal suspension~\cite{Huetter99,Graule94,Graule95,Mallamace06,Loewen98,Hecht06b,Linse00}.

However, on the mesoscopic length scale, often many different effects are in
a subtle interplay which makes it difficult to give quantitative predictions.
Computer simulations can help to study these systems in detail, e.g., to see the 
response of the system on a change in the particles' interactions.
Such a change can be achieved by adding salt to a suspension or by changing the
\pHvalue{}~\cite{Graule94,Graule95,Hecht06b,Hecht06}. However, in mesoscopic systems 
where many different effects contribute
to the overall behavior the parameter space is quite large and one has to 
perform a large number of different simulations, each of them requiring a lot of 
computing resources, to gain an understanding of all inter-relationships. 

Therefore, it is useful to be able to steer the simulation on-line. Steering in 
this context can mean to change the interaction potentials or other parameters like
an externally applied shear rate, but it can also mean to go back in the simulation
time and follow a different path starting from an earlier configuration, both 
induced by interaction with the user~\cite{Prins99,Chin03}.

For our study we have modeled an aqueous suspension of \Alumina-particles. The 
particles in our simulations are monodisperse spheres of $0.37\microns$ in diameter.
They interact via DLVO (Derjaguin, Landau, Vervey, Overbeek) potentials~\cite{DLVO,DLVO2} 
as well as a repulsive force ensuring excluded volume for the particles. The properties
of this model system have been investigated by simulations and experiments in 
previous works~\cite{Hecht05,Hecht06,Hecht06b,Hecht07,Oberacker01,Richter03}.
In this paper we discuss how steering of a simulation can help to explore the
parameter space and which problems may occur when using steering techniques.

In the following section we describe our simulation method and focus on the
implementation of the steering. Then, we discuss some examples of simulations 
where steering helps to explore the parameter space, but we also give examples 
in which a steered simulation might lead to different results if compared to an un-steered 
simulation. Finally, we summarize our results and draw a conclusion.

\section{Simulation Method}
Our simulation method is described in detail in \Refs{Hecht05,Hecht06,Hecht07} and
consists of two parts: a Molecular Dynamics (MD) code, which treats the
colloidal particles, and a Stochastic Rotation Dynamics (SRD) simulation
for the fluid solvent. In the MD part we include effective electrostatic
interactions and van der Waals attraction, known as DLVO
potentials~\cite{DLVO,DLVO2}. The repulsive term results from the 
surface charge of the suspended particles
\bea &&
  V_{\mathrm{Coul}} =
  \pi \epsilon_r \epsilon_0
  \left[ \frac{2+\kappa d}{1+\kappa d}\cdot\frac{4 k_{\mathrm{B}} T}{z e} 
    \right. \nonumber \\ && \;\left.  
        \tanh\left( \frac{z e \zeta}{4 k_{\mathrm{B}} T} \right)
  \right]^2 \times \frac{d^2 \mathrm{e}^{ - \kappa [r - d]}}{r},
 \label{eq_VCoul}
\eea
where $d$ denotes the particle diameter, $r$ the distance between the
particle centers, $e$ the elementary charge, $T$ the temperature,
$k_{\mathrm{B}}$ the Boltzmann constant, and $z$ is the valency of the
ions of added salt. $\epsilon_0$ is the permittivity of the vacuum,
$\epsilon_r=81$ the relative dielectric constant of the solvent, $\kappa$
the inverse Debye length defined by $\kappa^2 = 8\pi\ell_BI$, with
ionic strength $I$ and Bjerrum length $\ell_B = 7\angst$.  
We have related the effective surface potential $\zeta$ to the \pHvalue{} and 
the ionic strength of the solvent by means of a charge regulation model
in our previous work~\cite{Hecht06}. Thus, the particle interaction potentials
can be related to distinct experimental conditions. 
The second therm of the DLVO potentials which does not depend on the \pHvalue{} or 
the ionic strength is the attractive van der Waals interaction
($A_{\mathrm{H}}=4.76\cdot 10^{-20}\,\mathrm{J}$ is the Hamaker constant)~\cite{Huetter99}
\be
  V_{\mathrm{VdW}}  = 
     - \frac{A_{\mathrm{H}}}{12}  
     \left[ \frac{d^2}{r^2 - d^2} + \frac{d^2}{r^2} 
    + 2 \ln\left(\frac{r^2 - d^2}{r^2}\right) \right].
  \label{eq_VdW}
\ee
The attractive contribution $V_{\mathrm{VdW}}$ competes with the repulsive term and is responsible 
for the cluster formation one can observe for conditions in which the attraction dominates. 

Since DLVO theory is based on the assumption of large particle separations, it
does not correctly reproduce the primary minimum in the potential, which should appear 
at particle contact. Therefore, we cut off the DLVO potentials and model the minimum 
by a parabola as described in \Refs{Hecht05,Hecht07}. To ensure excluded
volume of the particles we use a repulsive (Hertzian) potential. Below the resolution 
of the SRD algorithm short range hydrodynamics is corrected by a lubrication force 
within the MD framework as explained in \Refs{Hecht05,Hecht06,Hecht07}.

For the simulation of a fluid solvent, many different simulation methods have been proposed: direct Navier Stokes solvers~\cite{Schwarzer96,Schwarzer97b,Schwarzer98,Schwarzer95}, 
Stokesian Dynamics (SD)~\cite{Brady88,Brady93,Brady96}, 
Accelerated Stokesian Dynamics (ASD)~\cite{Brady01,Brady04}, 
pair drag simulations~\cite{Silbert97a}, Brownian Dynamics (BD)~\cite{Huetter99,Huetter00}, 
Lattice Boltzmann (LB)~\cite{Ladd94,Ladd94b,Ladd01,Harting04}, 
and Stochastic Rotation Dynamics (SRD)~\cite{Inoue02,Padding04,Hecht05}.
These mesoscopic fluid simulation methods have in common that they make certain 
approximations to reduce the computational effort. Some of them include thermal noise intrinsically, or it can be included consistently. They scale differently with the 
number of embedded particles and the complexity of the algorithm differs largely.
In particular, there are big differences in the concepts how to couple the suspended 
particles to the surrounding fluid.

We apply the Stochastic Rotation Dynamics method (SRD) introduced by Malevanets and Kapral~\cite{Malev99,Malev00}. It intrinsically contains fluctuations, is easy to implement, and
has been shown to be well suitable for simulations of colloidal and polymer suspensions~\cite{Inoue02,Padding04,Gompper04,Gompper04b,yeomans-2004-ali,Hecht05,Hecht06}.
The method is also known as ``Real-coded Lattice Gas''~\cite{Inoue02} or 
as ``Multi-Particle-Collision Dynamics'' (MPCD)~\cite{Gompper05,Gompper06}.
It is based on so-called fluid particles with continuous positions and velocities. A streaming
step and an interaction step are performed alternately. In the streaming step, each particle $i$ 
is moved according to
\be
\label{eq_move}
\vec{r}_i(t+\tau)=\vec{r}_i(t)+\tau\;\vec{v}_i(t),
\ee
where $\vec{r}_i(t)$ denotes the position of the particle $i$ at time $t$ and $\tau$ is the 
time step.
In the interaction step the fluid particles are sorted into cubic cells of a regular 
lattice and only the particles within the same cell interact with each other
according to an artificial interaction. The interaction step is designed to exchange
momentum among the particles, but at the same time to conserve total energy and total 
momentum within each cell, and to be very simple, i.e., 
computationally cheap. Each cell $j$ is treated independently: first, the mean velocity 
$\vec{u}_j(t')=\frac{1}{N_j(t')}\sum^{N_j(t')}_{i=1} \vec{v}_i(t)$ in cell $j$ is calculated.
$N_j(t')$ is the number of fluid particles contained in cell $j$ at time $t'=t+\tau$.
Then, the velocities of each fluid particle in this cell are rotated according to
\be
\label{eq_rotate}
\vec{v}_i(t+\tau) = \vec{u}_j(t')+\vec{\Omega}_j(t') \cdot [\vec{v}_i(t)-\vec{u}_j(t')].
\ee
$\vec{\Omega}_j(t')$ is a rotation matrix, which is independently chosen at random
for each time step and each cell. We use rotations about one of the coordinate axes by 
an angle $\pm\alpha$, with $\alpha$ fixed.
The coordinate axis as well as the sign of the rotation are chosen at random, 
resulting in 6 possible rotation matrices. However, there is a great freedom to choose 
the rotation matrices. Any set of rotation matrices satisfying the detailed balance 
for the space of velocity vectors could be used here.
To remove anomalies introduced by the 
regular grid, one can either choose the mean free path to be sufficiently large 
or shift the whole grid by a random vector once per SRD time step~\cite{Ihle02a,Ihle02b}.

To couple the SRD and the MD simulation, basically three different methods have been introduced in the literature. Inoue~\etal~proposed a way to implement no slip boundary 
conditions on the particle surface~\cite{Inoue02}. To achieve full slip boundary 
conditions, Lennard-Jones potentials can be applied for the interaction between the 
fluid particles and the colloidal particles~\cite{Malev00, Padding06}. A more coarse
grained method was originally designed to couple the monomers of a polymer chain to 
the fluid~\cite{Malev00b, Falck04}, but in our previous work~\cite{Hecht05, Hecht06, Hecht06b} we have demonstrated that it can also be applied to colloids, as long as 
no detailed spacial resolution of the hydrodynamics is required. We use this coupling
method in our simulations and describe it shortly in the following. 

For the coupling of the colloidal particles to the fluid, they are sorted into the SRD 
cells and included in the SRD interaction step. The stochastic rotation is performed 
in the momentum space instead of the velocity space to take into account the difference 
of inertia between light fluid and heavy colloid particles.
We have described the simulation method in more detail in~\Refs{Hecht05,Hecht06,Hecht07}.

In the present work we report on several studies about pressure filtration, 
cluster formation in a steered simulation, and sedimentation. All these studies 
are proofs of principle, and therefore small simulations are performed. Typical 
parameters for these simulations are: 60 fluid particles per box, with $0.296\,\mu$m
extension in each direction. The volume of the box is set to be the same as the 
volume of a colloidal particle with diameter $0.37\,\mu$m. The system size is usually
30 to 60 boxes in each dimension. By matching the diffusion constant, density and
viscosity to a real suspension\cite{Hecht06}, the scaling scheme we have presented 
in \Ref{Hecht05} yields a scaled temperature in the simulation of 14\,mK and 
a scaled viscosity of $4.78\cdot10^{-8}$\,Pa\,s. To preserve the correct 
dynamics, characterized by the dimensionless numbers (Re, Pe, Kn,\dots) one 
has to rescale the potentials and all driving forces in the MD scheme by the 
same scaling factor, as well.

Let us now shortly sketch the technical realization of the steering interface in our 
simulation code. The program is an object oriented code written in C++. 
Each object contains virtual routines \verb|save()| and \verb|load()|, which write the
whole object data to or load it from a buffer, respectively. The buffer contains a 
plain text description of all variables contained in the object with their values, 
similar to the C++ source code one would write to initialize an instance of the 
class. The whole work flow of a simulation, including the actual MD loop as well as
data input/output tasks are described in this manner using specialized \verb|workstep|-classes\footnote{The workstep concept originally was introduced by M. Strau{\ss} in his SRD code~\cite{Ihle03c}.}%
, each providing a specialized \verb|work()|-routine which 
performs different actions depending on the actual class type of the respective object.
One of the \verb|workstep|-classes is designed to change a specified object by using its \verb|save()| and \verb|load()| routine. First, the current values are stored to a temporary buffer, then one or several variables may be overwritten by new values, and finally the object
to be modified is loaded again from the temporary buffer. The description of the changes may be 
read from standard input or from a file, similar to the simulation
setup, which is read during the initialization of any simulation. \verb|workstep|-class 
objects can also be included into the work flow at later times, or 
may even be disconnected from the usual work flow and bound to standard UNIX system 
signals. By default a workstep writing particle positions
to standard output and a workstep to change objects getting its buffer from
standard input are bound to the \verb|SIGTTOU| (``terminal output'') and \verb|SIGTTIN| (``terminal input'') signal, respectively.
This allows to embed the simulation program into a framework of shell scripts which 
generate the appropriate input, redirect the output to a visualization tool and 
send the signals according to the user interaction. This can be realized in a 
client-server fashion, even on different hosts and platforms using appropriate 
scripts and TCP/IP connections. 

\section{Results}

After the more technical paragraph in the previous section we now turn to a discussion 
of the advantages of such a steering approach for computer simulations and highlight some 
pitfalls resulting from the physical background in the context of steering. This section is 
subdivided into several subsections each of them focusing on a particular example to illustrate
the steering approach in practice.

\subsection{Pressure filtration}

\begin{figure}
\bc\mbox{\epsfig{file=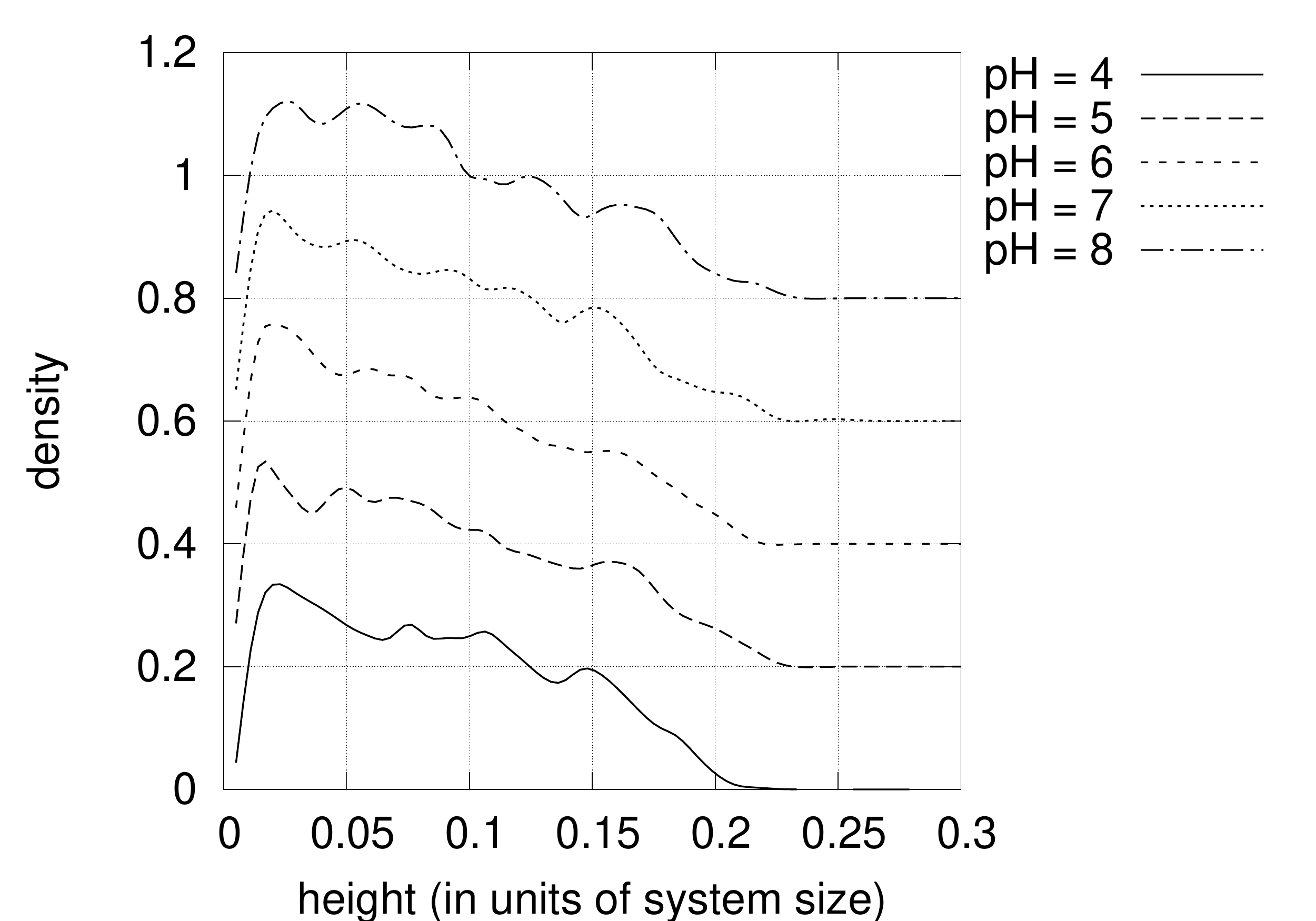,width=\epswidth}}\ec
\caption{Local density expressed in terms of volume fraction of the solid particles constituting 
the filter cake depending on the \pHvalue{}. The ionic strength is kept constant at $I=3\,$mmol/l, 
and the volume fraction at $\Phi=5\%$. 
The plots for different \pHvalue{}s are shifted against each other vertically by $0.2$ for 
better visibility. The shape of the profile differs, since the spacing of the particles 
depends on their interactions.}
\label{fig_SediHeight}
\end{figure}

In this case a filtration of a suspension is simulated. The suspended particles cannot 
pass the filter, whereas the fluid passes through the filter without resistance in an idealized
filtration process. The suspended particles agglomerate in front of the filter and 
form a so-called filter cake. Since the dynamics of the particles is not only governed
by the hydrodynamics of the fluid, but also by their DLVO interaction, the density
and the structure of the filter cake will depend on the \pHvalue{} and the ionic
strength.

In our simulation we drive the fluid by an applied force pointing downwards acting 
in a small region close to the upper boundary of the system. For the fluid we apply 
fully periodic boundary conditions, whereas the boundaries for the suspended particles are 
closed in $z$-direction. In this way, the fluid is forced to stream in vertical direction and to 
drag the particles to the bottom of the system. In test simulations we have chosen constant
ionic strength $I=3\mmoll$ and a constant overall volume fraction of $\Phi=5\%$, and 
simulated the filtration process for several \pHvalue{}s.

Depending on the interactions, different internal structures of the filter cake are formed 
when the density of the particles increases at the bottom of the system. To gain a good statistics 
for the structure large simulations are needed. However, in a filtration process, usually low
initial densities are used, so that very large simulation volumes are needed to obtain 
large particle numbers for good statistics. On the other hand, large simulation 
volumes also imply large data files and long transient times until the filter cake 
is formed and a considerable part of the particles has reached the bottom of the system.

This is a typical problem, in which one would like to observe the running simulation occasionally
to check if the filter cake is already formed and when the structure does not change 
anymore. One would like to measure pressure profiles, local streaming 
velocities, or simply the final density profile of the filter cake. Here, steering 
would mean to initiate data acquisition by user interaction. This is of advantage if
an action, like starting to average certain data, should be initiated when conditions 
are fulfilled, which are difficult to check automatically. Since the density and structure 
of the filter cake is a priori unknown, an automated check is difficult to implement. 

Additionally, the upper boundary of the filter cake is diffuse and depends
on the conditions (\pHvalue{} and ionic strength) as well. In \fig{fig_SediHeight} 
the dependence of the density profiles of filter cakes obtained in simulations
for different \pHvalue{}s are shown. With increasing \pHvalue{} the height of the filter cake increases and the density profile changes slightly. Some voids in the filter cake diminish the 
local density if compared to a dense packing. At the bottom of the filter cake, layers 
of particles are present, whereas in the upper regions the structure becomes more 
irregular. However, the structure of the sediment and its height will also depend 
on the pressure exerted on the fluid and probably 
on the initial volume fraction. Larger simulations than this proof of principle 
are needed to be able to quantify the \pHvalue{}-dependence of the height of the
filter cake. However, simulations without driving force on the fluid, can also
help to understand this system. Therefore, we explore the parameter space in the absence
of a driving force in the following subsection.

\subsection{Observation of cluster formation}

\begin{figure*}
\bc\mbox{a)\epsfig{file=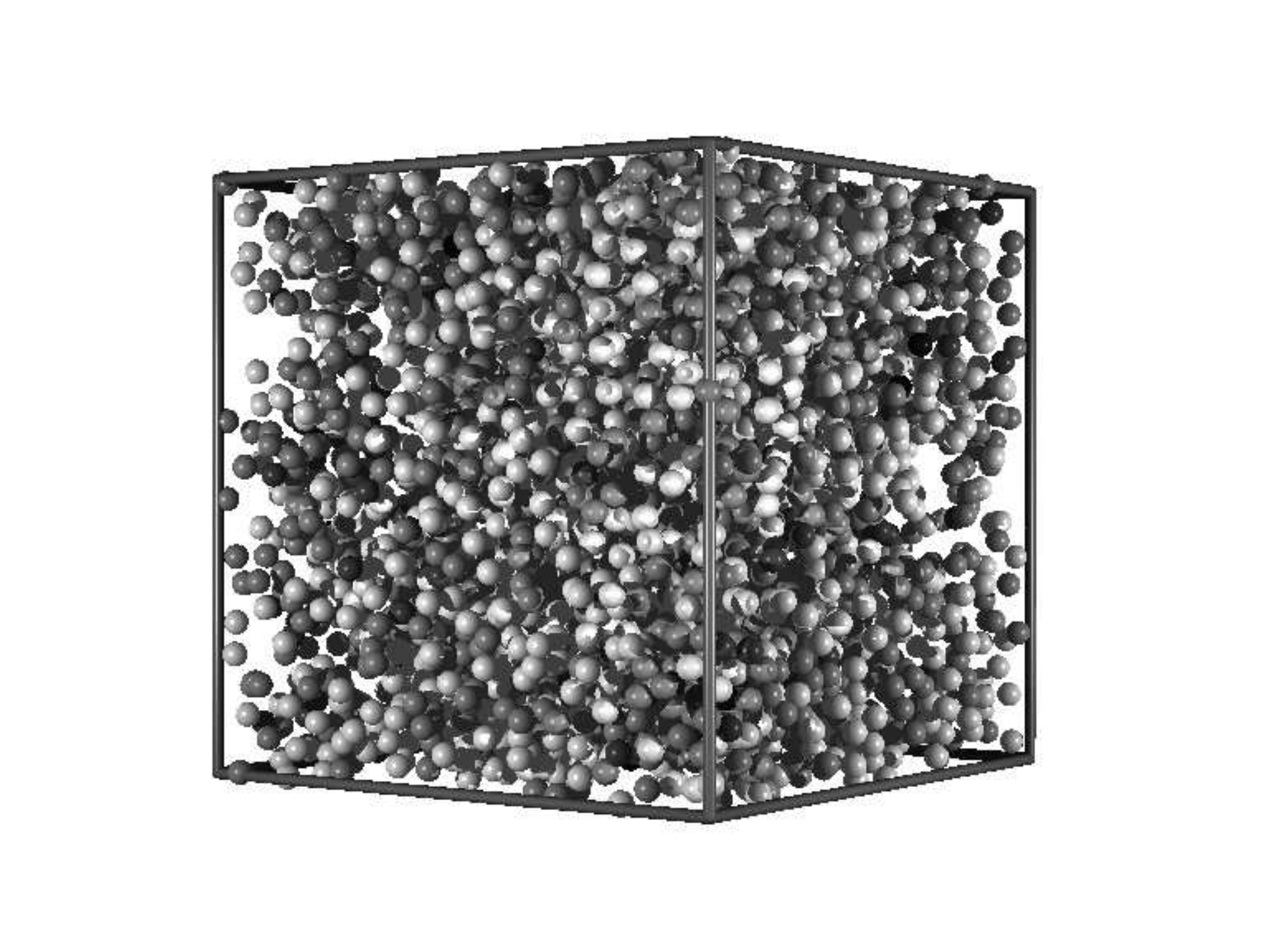,width=0.45\linewidth}}
\mbox{b)\epsfig{file=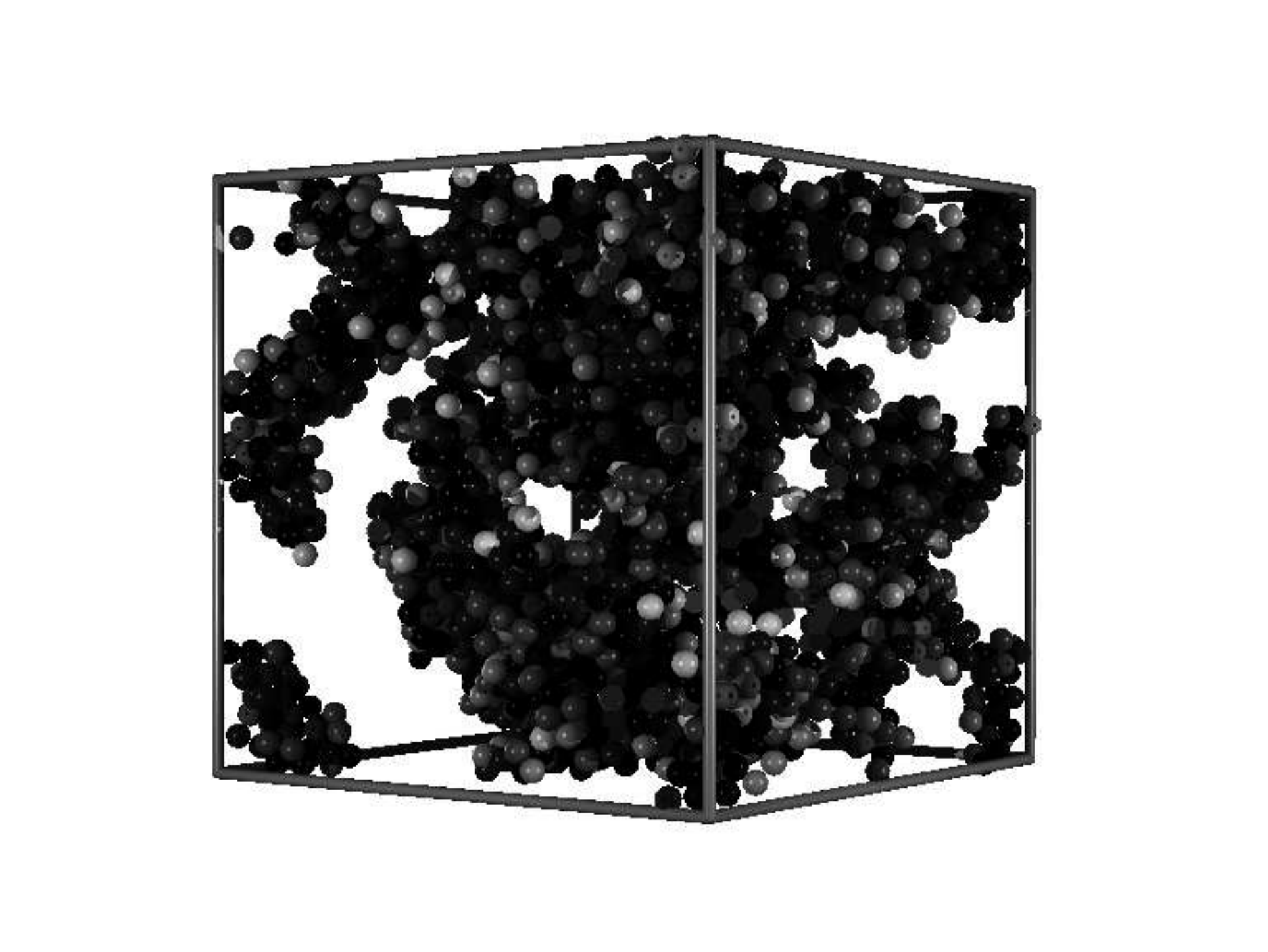,width=0.45\linewidth}}\ec
\bc\mbox{c)\epsfig{file=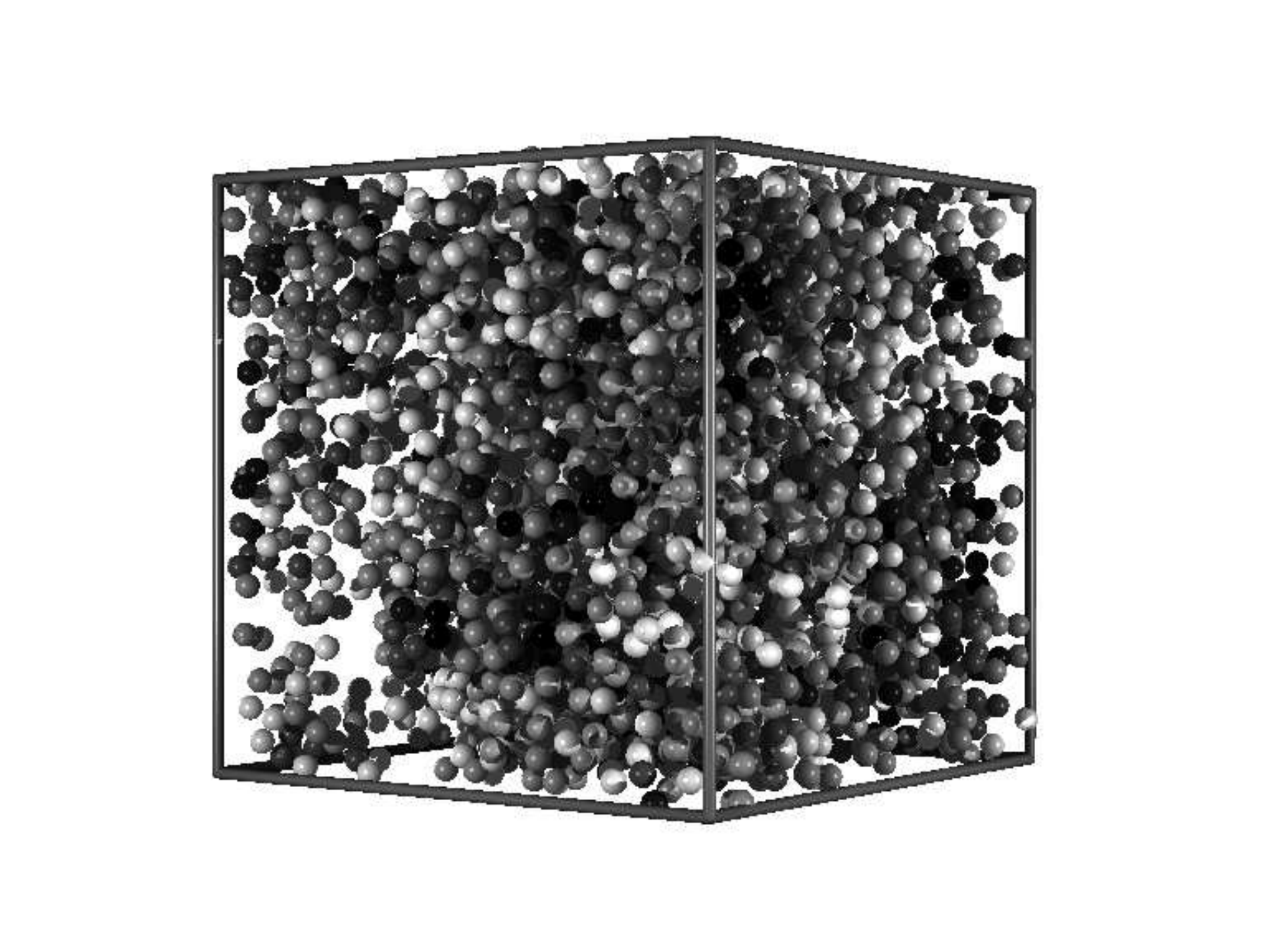,width=0.45\linewidth}}
\mbox{d)\epsfig{file=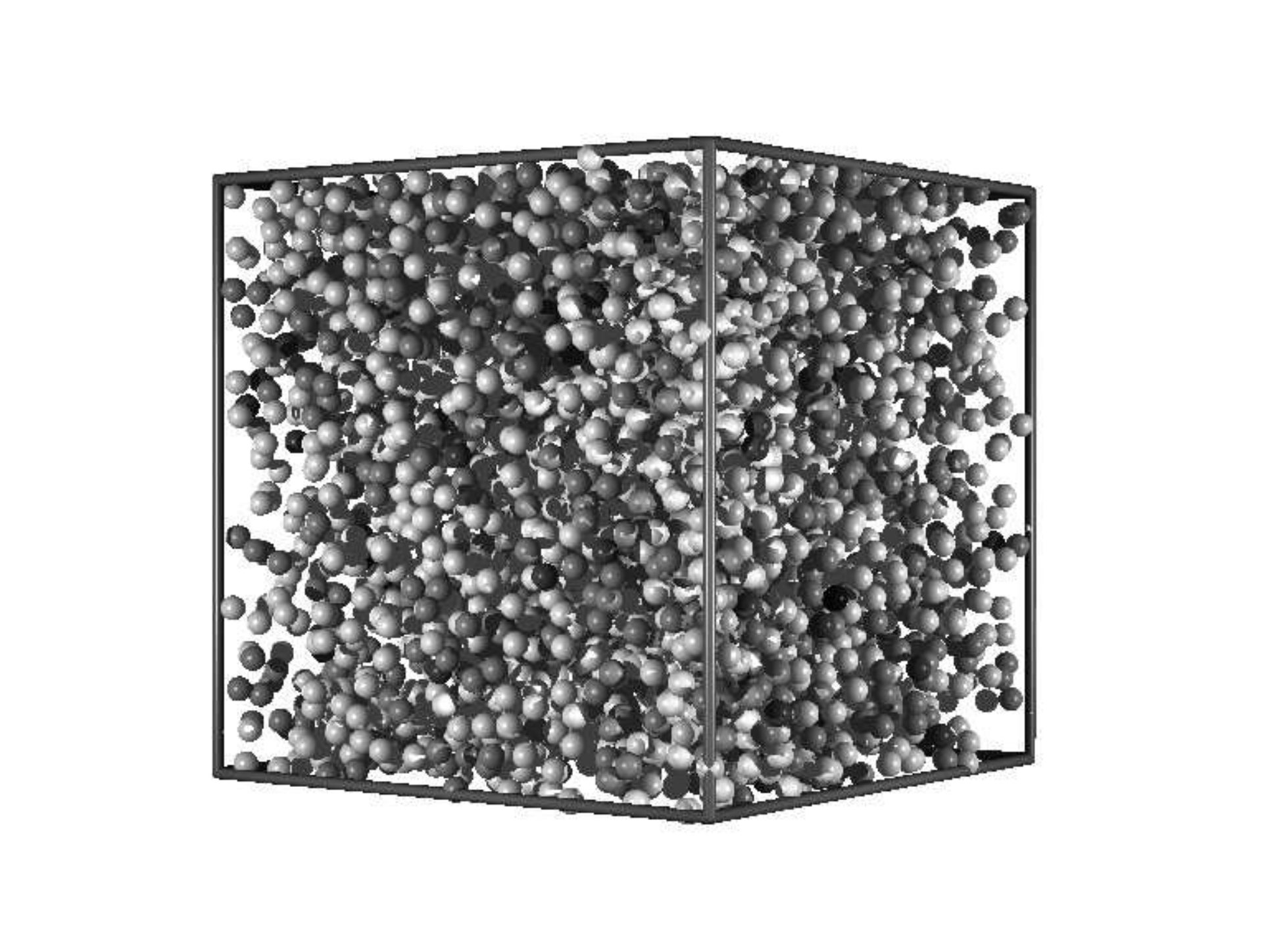,width=0.45\linewidth}}\ec
\caption{Memory effects in a steered simulation: the grayscale
denotes the coordination number. Particles closer than 2.4 radii are considered to
be neighbors. Dark particles in this sense are highly coordinated, bright ones have 
low coordination numbers. a) right after initialization, homogeneous distribution 
of the particles, b) cluster formation after tuning the potentials to be
attractive, c) still inhomogeneous regions after steering the simulation 
out of the clustered region, d) the system is homogeneous again when diffusion
has counterbalanced the density inhomogeneities. The steering path in the parameter
space as well as the points a)--d) thereon are illustrated in \fig{fig_steeringpath}.}
\label{fig_Steered}
\end{figure*}

Since several parameters, namely the volume fraction, \pHvalue{}, ionic strength, 
external driving forces influence the system, a first step to understand its behavior is 
to examine it in the absence of external forces. Still, two parameters--%
the \pHvalue{} and the ionic strength--govern the dynamics of the system. If the 
potentials are attractive, cluster formation can be observed~\cite{Hecht06c}, 
provided the attraction is strong enough compared to thermal fluctuations. In our 
previous work~\cite{Hecht06b,Hecht07} we have explored this part of the parameter 
space for some distinct volume fractions by running numerous individual simulations. 
However, with the newly implemented steering approach one can detect 
boundaries between different regions of the stability diagram faster and more sensitively. 

\begin{figure*}
\bc\mbox{\epsfig{file=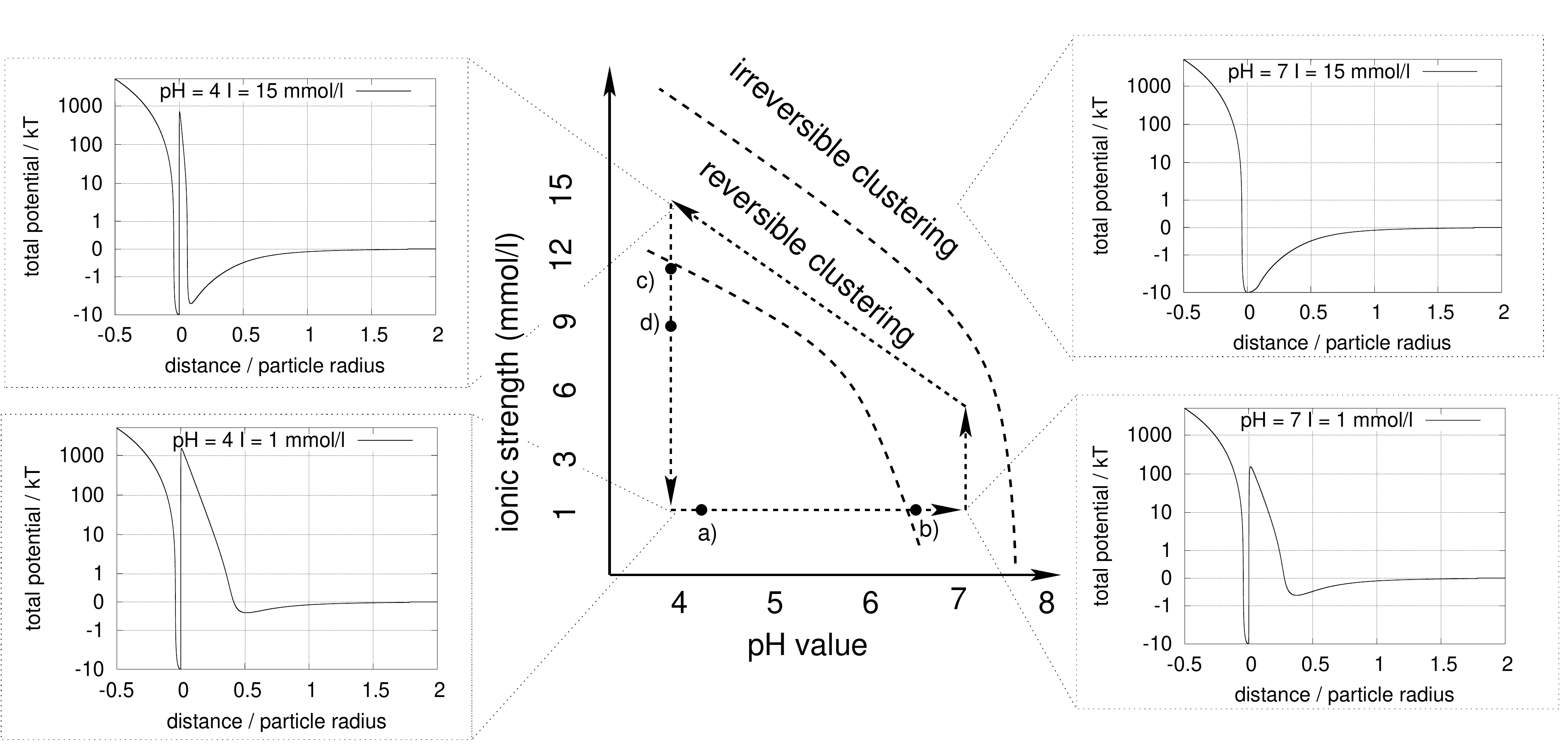,width=\linewidth}}\ec
\caption{Steering path of the simulation: we start at the bottom left corner at low
\pHvalue{} and low ionic strength and steer the simulation along the dashed path through
parameter space. The region of reversible cluster formation is reached, but the 
barrier between primary and secondary minimum of the DLVO potentials prevents irreversible
clustering: when reaching the starting point again, all clusters have vanished.
The boundary of the clustering region is sketched: Reversible clustering is observed 
when the depth of secondary DLVO minimum is in the order of $\kB T$ and irreversible
clustering appears when the height of the barrier between the two DLVO minima is of the
order of $\kB T$ or less. The labeled black dots mark the positions at which the 
snapshots of \fig{fig_Steered} are taken. The insets show the total potential at some 
selected extreme cases (see text).}
\label{fig_steeringpath}
\end{figure*}

In \fig{fig_Steered} we illustrate the effects of changing
the properties of the interaction potentials. 
All the snapshots belong to the same simulation and represent different 
points of the trajectory the system is steered through parameter space: a) The 
initial situation with repulsive potentials, shortly after initialization of the 
system, b) after crossing the boundary to the clustered region, c) shortly after 
steering back to the suspended regime, where still some inhomogeneities in the 
system can be observed, and finally d) after some time when diffusion has restored
a homogeneous system. This time depends on the distances between the clusters and 
on the characteristic diffusion time of the particles. The steering path in the
configuration space is shown in \fig{fig_steeringpath}. The points at which the 
snapshots of \fig{fig_Steered} are taken, are marked by black dots. In the insets 
in \fig{fig_steeringpath} the total potential is plotted for several cases.
During the simulation we change the 
\pHvalue{} and the ionic strength and we drive the system along a closed trajectory 
in parameter space. We start at low ionic strength $I=1\mmoll$ and $\pH = 4$ (potential
shown) and first increase the \pHvalue{} (potential shown for $I=1\mmoll$ and $\pH = 7$), then increase the ionic strength, then reduce the 
\pHvalue{} again while further increasing the ionic strength (potential shown for $I=15\mmoll$ and $\pH = 4$) and finally decrease 
the ionic strength to return to the starting point. We choose the trajectory such 
that the barrier between the primary and the more shallow, secondary minimum of the DLVO 
potential does not go below 5 $\kB T$. Since the secondary minimum becomes deeper during 
the simulation cluster formation can be observed. However, the barrier between the 
primary and the secondary minimum of the DLVO potentials prevents irreversible 
aggregation in the primary minimum. A case in which the barrier vanishes and irreversible 
aggregation would appear is also shown in the upper right inset 
($I=15\mmoll$ and $\pH = 7$).
In any case, in our simulation all clusters dissolve again after steering back to 
a lower \pHvalue{} and lower ionic strength (see \fig{fig_Steered}\,d)). This confirms 
that we have reversible clustering in the secondary minimum of the DLVO potentials. 

One of the advantages of a steered simulation is, that one can more accurately 
detect the onset of the cluster formation than by starting individual simulations
for different conditions. One might predict the boundary by evaluating the depth of
the secondary minimum in the potential, but since also the hydrodynamic damping 
forces and an eventually applied driving pressure influence this process the prediction
of the boundary can become more complicated. We have experienced similar difficulties
in the context of shear flow simulations of the same colloidal 
suspension\,\cite{Hecht06,Hecht06b}.

A quantity to support the visualization of clusters is the coordination number.
We consider two particles whose centers are closer than $2.4$ particle radii as neighbors and 
define the coordination number of each particle as the number of its neighbors in the sense 
just described. The particles in \fig{fig_Steered} are drawn in a grayscale corresponding 
to their coordination  number, where dark particles are highly coordinated particles and 
bright ones are those with a low coordination number. At the beginning of the simulation
(top left) the particles are distributed homogeneously over the whole simulation volume.
After steering the simulation in the parameter space into the region of attractive potentials, 
cluster formation can be observed (top right). When the potentials are made less attractive 
again, the clusters dissolve (bottom left). However, the particles preferably stay at their
positions and thus, the inhomogeneities in the system do not disappear immediately,
which can be seen also in the dark color denoting high coordination numbers. 
After some time, diffusion restores the homogeneity of the system (bottom right).
However, if the \pHvalue{} or the ionic strength is increased too much, so that the 
barrier between the primary  and the secondary minimum of the DLVO potentials vanishes, 
the clustering process becomes irreversible, which we avoided by the choice of the
steering path (\fig{fig_steeringpath}).

In this example the simulation remembers the interaction by the user. Even if the steering 
path is selected carefully inside the reversible range, the simulated suspension needs a 
characteristic time to relax after changing the interactions. The interaction by the user
can be seen as a perturbation in the physical sense and the system needs time to adopt to the
new situation. In these cases special care is advised when steering a simulation.
We illustrate this by another example in the following subsection.

\subsection{sedimentation: hydrodynamic interaction}

\begin{figure}
\bc\mbox{\epsfig{file=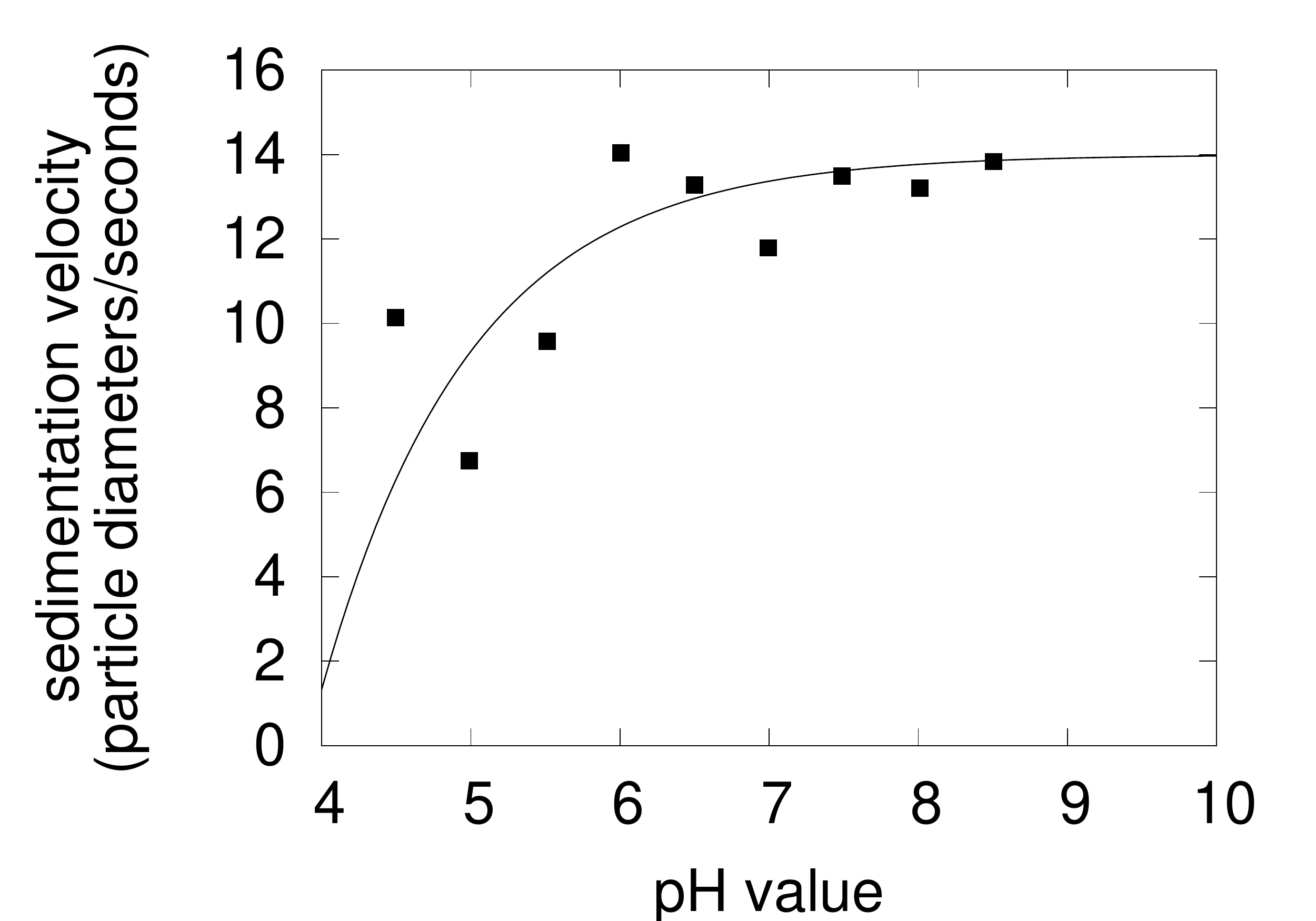,width=\epswidth}}\ec
\caption{Sedimentation velocity of a $\Phi=5\%$ suspension in a closed vessel.
The sedimentation velocity of the particles in the upper part of the system is 
averaged over several time steps. The particles at the bottom of the system are
not encountered. The velocity depends on the particle-particle interaction. Clusters
settle down faster than individual particles. Symbols denote simulation results,
the line is a guide to the eye.}
\label{fig_VSedi}
\end{figure}

Not only the particle positions and velocities, but also hydrodynamic interactions
can influence the behavior of the system.
Therefore, inhomogeneous particle distributions induced by steering the simulation first 
into a clustered region of the parameter space before simulating a different situation,
might influence the result: we show in the following paragraph that a sedimentation process
can happen faster, if cluster formation by attractive interactions creates density 
inhomogeneities in the system. 

The friction force a single particle feels in a sedimentation process can be calculated 
analytically, but it differs if other particles are present. In \Ref{Hecht05} we have
confirmed that the sedimentation velocity in our simulation depends on the volume fraction, as it is well-known from sedimentation theory~\cite{RichardsonZaki}. 
But even at constant volume fraction the sedimentation velocity can be different in 
different configurations. 
If the particles form clusters, the whole cluster settles down and the fluid streams around 
the whole cluster. The resistance is much less compared to the case when the particles 
are distributed homogeneously and the fluid streams around each of them separately
(compare \Ref{Hecht07} and references therein). 
In \fig{fig_VSedi} we have plotted the sedimentation velocity evaluated in several
simulations, which only differ by the potentials. The volume fraction is the same 
for all of them. However, as one can see in the figure, the sedimentation 
velocity is smaller for low \pHvalue{}s. 

This can be explained as follows: Since for increasing \pHvalue{} the 
potentials become attractive, the particles form clusters and settle down faster. 
This effect is purely due to the hydrodynamics of the system. Since the streaming
field depends on all particle positions and velocities not only at a given time,
but also on their history, steering in this context may be very dangerous. 
When long range interactions, like hydrodynamics, which also depends on the 
history of the system, is important, steering should be avoided.
In this case, to obtain reasonable results, one has to start again from an initial
configuration when changing the interactions. 

\section{Conclusion}

Bearing in mind the underlying physics, one can roughly estimate the limits 
of steered simulations. Let us summarize the points we have brought forward in this paper.
First of all, steering can save computing time when aiming at a rough understanding of the 
influences on a system, especially when looking for the ``interesting points'' in parameter
space, e.g., when searching for transition lines between different phases. 
We have shown this by the onset of cluster formation due to modifying the interaction
potentials in a model suspension. Steering can be used as well to start data acquisition after a transient at the beginning 
of a simulation or to adjust the frequency for the data acquisition according to the current
state of the simulation. This is especially interesting for large simulations with long 
transient times, as illustrated by the example of filter flow. We also explained why in some 
cases steering may be seen as a 
perturbation of the system, especially when the interaction potentials are changed. 
Special care is advised when dealing with non ergodic systems, memory effects, or long range
interactions, like the role of hydrodynamics in sedimentation in our last example.

\section*{Acknowledgments}
\noindent
We thank H.J.~Herrmann for valuable collaboration and his support. The High
Performance Computing Center Stuttgart, the Scientific Supercomputing Center
Karlsruhe and the Neumann Institute for Computing in J\"ulich are highly
acknowledged for providing the computing time and the technical support needed
for our research. J.H. thanks the DFG for financial support within the priority 
program ``nano- and microfluidics''. M.H. thanks the ICMMES for travel 
funds and the DFG for financial support within SFB 716.


\end{document}